\documentclass[twoside,twocolumn,9pt]{article}
\usepackage{extsizes}
\usepackage[super,sort&compress,comma]{natbib} 
\usepackage[version=3]{mhchem}
\usepackage[left=1.5cm, right=1.5cm, top=1.785cm, bottom=2.0cm]{geometry}
\usepackage{balance}
\usepackage{mathptmx}
\usepackage{sectsty}
\usepackage{graphicx} 
\usepackage{lastpage}
\usepackage[format=plain,justification=justified,singlelinecheck=false,font={stretch=1.125,small,sf},labelfont=bf,labelsep=space]{caption}
\usepackage{float}
\usepackage{fancyhdr}
\usepackage{fnpos}
\usepackage[english]{babel}
\addto{\captionsenglish}{%
  
}
\usepackage{array}
\usepackage{droidsans}
\usepackage{charter}
\usepackage[usenames,dvipsnames]{xcolor}
\usepackage{setspace}
\usepackage[compact]{titlesec}
\usepackage[bookmarks=false,colorlinks]{hyperref}
\hypersetup{linkcolor=magenta,citecolor=MidnightBlue,filecolor=Plum,urlcolor=MidnightBlue}

\usepackage[normalem]{ulem}

\definecolor{cream}{RGB}{222,217,201}

\usepackage{booktabs,microtype}

\begin{document}

\pagestyle{fancy}
\thispagestyle{plain}
\fancypagestyle{plain}{
\renewcommand{\headrulewidth}{0pt}
}

\makeFNbottom
\makeatletter
\renewcommand\LARGE{\@setfontsize\LARGE{15pt}{17}}
\renewcommand\Large{\@setfontsize\Large{12pt}{14}}
\renewcommand\large{\@setfontsize\large{10pt}{12}}
\renewcommand\footnotesize{\@setfontsize\footnotesize{7pt}{10}}
\makeatother

\renewcommand{\thefootnote}{\fnsymbol{footnote}}
\renewcommand\footnoterule{\vspace*{1pt}%
\color{cream}\hrule width 3.5in height 0.4pt \color{black}\vspace*{5pt}} 
\setcounter{secnumdepth}{5}

\makeatletter 
\renewcommand\@biblabel[1]{#1}            
\renewcommand\@makefntext[1]%
{\noindent\makebox[0pt][r]{\@thefnmark\,}#1}
\makeatother 
\renewcommand{\figurename}{\small{Fig.}~}
\sectionfont{\sffamily\Large}
\subsectionfont{\normalsize}
\subsubsectionfont{\bf}
\setstretch{1.125} 
\setlength{\skip\footins}{0.8cm}
\setlength{\footnotesep}{0.25cm}
\setlength{\jot}{10pt}
\titlespacing*{\section}{0pt}{4pt}{4pt}
\titlespacing*{\subsection}{0pt}{15pt}{1pt}

\fancyfoot{}
\fancyfoot[LO,RE]{\vspace{-7.1pt}\includegraphics[height=9pt]{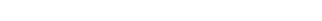}}
\fancyfoot[CO]{\vspace{-7.1pt}\hspace{13.2cm}\includegraphics{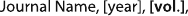}}
\fancyfoot[CE]{\vspace{-7.2pt}\hspace{-14.2cm}\includegraphics{head_foot/RF}}
\fancyfoot[RO]{\footnotesize{\sffamily{1--\pageref{LastPage} ~\textbar  \hspace{2pt}\thepage}}}
\fancyfoot[LE]{\footnotesize{\sffamily{\thepage~\textbar\hspace{3.45cm} 1--\pageref{LastPage}}}}
\fancyhead{}
\renewcommand{\headrulewidth}{0pt} 
\renewcommand{\footrulewidth}{0pt}
\setlength{\arrayrulewidth}{1pt}
\setlength{\columnsep}{6.5mm}
\setlength\bibsep{1pt}

\makeatletter 
\newlength{\figrulesep} 
\setlength{\figrulesep}{0.5\textfloatsep} 

\newcommand{\topfigrule}{\vspace*{-1pt}%
\noindent{\color{cream}\rule[-\figrulesep]{\columnwidth}{1.5pt}} }

\newcommand{\botfigrule}{\vspace*{-2pt}%
\noindent{\color{cream}\rule[\figrulesep]{\columnwidth}{1.5pt}} }

\newcommand{\dblfigrule}{\vspace*{-1pt}%
\noindent{\color{cream}\rule[-\figrulesep]{\textwidth}{1.5pt}} }

\makeatother

\twocolumn[
  \begin{@twocolumnfalse}
{\includegraphics[height=30pt]{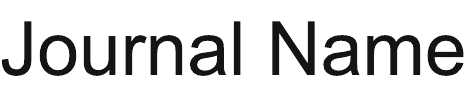}\hfill\raisebox{0pt}[0pt][0pt]{\includegraphics[height=55pt]{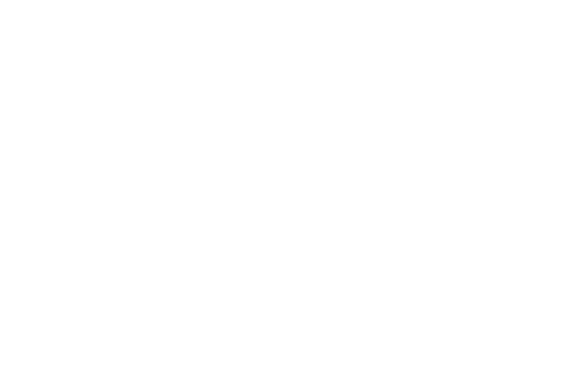}}\\[1ex]
\includegraphics[width=18.5cm]{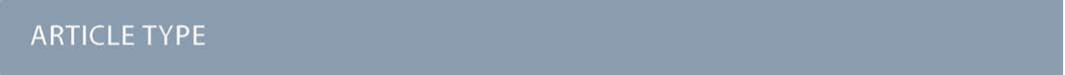}}\par
\vspace{1em}
\sffamily
\begin{tabular}{m{4.5cm} p{13.5cm} }

\includegraphics{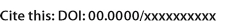} & \noindent\LARGE{\textbf{Instability-Avoiding Active Learning for Cluster Expansions in Complex Multielement Materials$^\dag$}} \\
\vspace{0.3cm} & \vspace{0.3cm} \\

 & \noindent\large{Michael J.\ Waters,\textit{$^{a}$} and James M.\ Rondinelli$^{\ast}$\textit{$^{a}$}} \\

\includegraphics{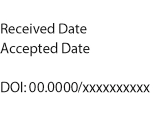} & \noindent\normalsize{The high efficiency of cluster expansions make them appealing for studying chemical disorder in complex composition spaces such as in multi-principal element alloys (MPEAs).
Several works have attempted to address the rapidly growing training cost with number of chemical species through active learning, transfer learning, and chemical embedding.
However, many composition spaces have large regions where the host lattice becomes dynamical unstable, which are often avoided \textit{a priori} so as to not generate expensive but inapplicable training data. 
Here, we demonstrate a procedure for integrating stability classification within an active learning workflow to autonomously avoid calculations for unstable structures.
Our workflow augments the stability classification procedure with Mahalanobis distance-based structure selection to ensure model robustness by training set diversification.
We benchmark our methods by training a cluster expansion for the complex FCC MPEA spanning the Ni-Fe-Cr-Al-Ti-Si alloy space, in which only Ni and Al are thermodynamically stable as FCC.}\\
\end{tabular}

 \end{@twocolumnfalse} \vspace{0.6cm}

  ]

\renewcommand*\rmdefault{bch}\normalfont\upshape
\rmfamily
\section*{}
\vspace{-1cm}


\footnotetext{\textit{$^{a}$~ Department Materials Science and Engineering, Northwestern University, Evanston, Illinois, USA. Tel: +1 847-491-3198; E-mail: jrondinelli@northwestern.edu}}

\section{Introduction}

Multi-principal element alloys (MPEAs) are promising for high-temperature \cite{HEAs_for_hightemp}, low-temperature \cite{HEAs_for_lowtemp}, nuclear \cite{HEAs_for_nuclear}, functional  \cite{HEAs_functional_materials}, and even catalytic applications \cite{mpea_ammonia}. 
In such cases, their performance is tied to properties that are modulated by chemical short range ordering (SRO) at the atomic scale such as: ductility \cite{SRO_and_mechanical_properties}, stacking fault energies \cite{SRO_modulates_stacking_fault_energies}, fatigue life \cite{SRO_enhances_fatigue_life} radiation resistance \cite{SRO_enhances_radiation_resistance}, and magnetism \cite{Magnetism_and_SRO}.
Motivated by the growing evidence that SRO also affects aqueous corrosion processes, we focus here on the relationship between SRO and passivation. 
Previously, Xie \textit{et al.} showed that in Fe/Ni-Cr alloys, Cr site percolation, which is modulated by SRO, is required for the formation of a protective passivating oxide \cite{Xie_Sieradzki_2021,roy2024effect}. 
In MPEAs, however, multiple passivating species may be present, each with distinct effectiveness and complex interactions that influence both SRO and the concentration-dependent percolation thresholds.
To investigate SRO-passivation behavior in a multi-passivator MPEA, we selected the FCC Fe-Ni-Cr-Al-Ti-Si composition space.
This system expands the Fe-Ni-Cr base of austenitic stainless steels by incorporating  Al, Ti, and Si as additional earth-abundant passivating elements, while retaining the FCC parent lattice.
The chemical and configurational complexity of such MPEAs presents significant challenges for computational materials science. 
Both the number of distinct chemical interaction permutations and the number of uniform-composition grid points increase rapidly with number of alloying elements considered (Section 1 of the Supporting Information, SI). 
To enable a tractable search of equilibrium ordering and SRO across this space, we use the established strategy of fitting a cluster expansion (CE) model to \textit{ab initio} data, followed by Monte Carlo (MC) simulations  \cite{ReviewCE_MATSCI}.
In a CE, the configurational Hamiltonian is decomposed into contributions from $N$-body clusters (pairs, triplets, quadruplets, \textit{etc.}) of lattice sites, with each unique species decoration of a cluster contributing an independent interaction parameter.
The resulting cluster functions form a complete basis across both species and lattice decorations and serve as the  input vector for the CE model.
For computational tractability, the expansion is truncated in body order and, when formulated for real-space, by limiting the spatial extent of the cluster interactions.
Although CEs most commonly represent thermodynamic potentials, \textit{e.g}., the internal energy for MC simulations, the framework is general \cite{tensorial_cluster_expansion_vandeWalle2008}. 
Any intensive property can be expanded to capture its dependence on the underlying lattice decorations.
Examples include band gap \cite{bandgap_CE} and  tensorial  properties such as strain  \cite{tensorial_cluster_expansion_vandeWalle2008,CPS2,CPS3}. 

\begin{table}[b]
    \centering
    \caption{The number symmetrically inequivalent supercells (SIS) of the primitive FCC lattice for various orbit (supercell) sizes. Total $\mathrm{SIS} = 47,720$ for atoms in orbits 1--6}
    \begin{tabular}{lllllll}
    \toprule
        Atoms in orbit ($N$) & 1 & 2 & 3 & 4 & 5 & 6  \\
        Unique SIS ($N$) & 6 & 30 & 150 & 1,350 & 3,984 & 42,200  \\
        \bottomrule
    \end{tabular}
    \label{tab:table_enumeration_counts}
\end{table}

The key advantage of CE for sampling large MPEA composition spaces lies in their computational efficiency. 
Because they use a purely decoration-based Hamiltonian, millions of MC trials can be performed across thousands of compositions at relatively low cost. 
CE construction typically relies on \textit{ab initio} (usually DFT) relaxations on small supercells, representing  unique crystallographic orbit decorations. 
We refer to these as symmetrically inequivalent supercells (SIS).
Although alternative training-set generation schemes exist \cite{CLEASE,CELL,icetTutorial2024}, SIS-based enumeration remains attractive because the associated DFT calculations are comparatively less expensive.

Prior to CE fitting, however, relaxed structures must be screened.
Structures that relax far from the ideal lattice, which indicate dynamical instability rather than small equilibrium displacements, are excluded.
Additionally, structures lying far above the convex hull  are often discarded or given lower weight during fitting \cite{atat_automation}.
In a six-component FCC alloy space, the number of SIS grows rapidly with orbit size (\autoref{tab:table_enumeration_counts}, Table S1), making indiscriminate DFT evaluations impractical. 
Therefore, the first design goal of our workflow is to reduce the DFT-simulation burden by intelligently selecting the most informative training structures.

To this end, we employ an active learning strategy based on an ensemble of CE models.
The ensemble variance, \textit{i.e.}, the disagreement between the model predictions, is used as an uncertainty metric for ranking prospective structures.
Alternatively, prior works have demonstrated the effectiveness of Bayesian active learning and compressed sensing approaches for improving CE training \cite{AxelBayesian, mueller_exact_expressions_for_selection, MuellerBayesian, icetTutorial2024, compressive_sensing_Hart, Bayesian_compressive_sensing_Hart}.
While we rely on enumerated structures, candidate configurations may also be generated via simulated annealing or direct optimization for maximum model uncertainity or  orthogonality in cluster-function space 
\cite{icetTutorial2024,CLEASE,alloy_design_inverse_CE}.
Additional methodological details are available in  Refs.\  \cite{icetPaper2019,icetTutorial2024,MuellerBayesian,FC_Cluster_linear_erhart, AxelBayesian, CLEASE, mueller_exact_expressions_for_selection, ANOVA_Barroso-Luque_Cedar_2024}.  

\begin{table}[b]
    \centering
    \caption{Formation energies relative to the convex hull from OQMD~\cite{OQMD_2013,OQMD2015} (meV/atom). A value of 0 indicates the phase is on the convex hull (thermodynamically stable)}
    \begin{tabular}{lllll}
        \toprule
        Element & FCC & BCC & HCP & Diamond Cubic \\
        \midrule
        Fe & 161 & 0 & 86 & 1244 \\
        Ni & 0 & 91 & 23 & 1198 \\
        Cr & 394 & 0 & 406 & 2076 \\
        Al & 0 & 96 & 33 & 747 \\
        Ti & 70 & 114 & 14 & 2116 \\
        Si & 537 & 523 & 508 & 0 \\
        \bottomrule
    \end{tabular}

    \label{tab:table_phase_stabilities}
\end{table}

Active learning shows great efficiency in particular for well-behaved alloys, where the target lattice is dynamically stable across compositions and relevant intermetallics correspond to  ordered decorations of that lattice.
However, the Fe-Ni-Cr-Al-Ti-Si  space is poorly behaved as an FCC alloy system. 
Only Ni and Al are stable in FCC, while FCC Cr and Si are highly unfavorable (\autoref{tab:table_phase_stabilities}), and multiple non-FCC intermetallics such as the B2 NiAl and FeTi are present. 
Consequently, many actively selected SIS are likely to relax away from FCC or remain trapped in unphysically high-energy metastable states. 
Both outcomes lead to filtered structures and therefore wasteful DFT calculations. 
This motivates our second design goal: incorporating instability avoidance directly into candidate structure selection, with parity to post-DFT filtering, in a generalizable manner.
This can be critical to prevent some active learning workflows from selecting under-sampled, unstable structures with increasing bias in each iteration as their uncertainty metrics remain large relative to those of stable structures.
For energetic filtering, low-energy structures can be identified using the current CE iteration to predict hull proximity, an approach widely adopted in tools such as 
Alloy-Theoretic Automated Toolkit (ATAT) \cite{atat_automation,ATAT_user_guide} and CLEASE \cite{CLEASE}. 

Predicting lattice stability \emph{a priori} is more challenging. 
Here we introduce a second CE not of energy, but of the average atomic displacement from ideal FCC. 
This displacement model predicts the mean post-relaxation displacement, based on the assumption that average displacements grow as compositions or local ordering approach the limits of dynamical stability.
This approach is lattice agnostic and does not require any \textit{a posteriori} knowledge of stability regimes.
To further ensure robustness, actively selected structures are augmented with Mahalanobis distance biased sampling to promote data diversity.
Six-element composition spaces remain exceptionally large for cluster expansions trained on \textit{ab initio} data.
To our knowledge, only three prior works report \textit{ab initio}-based CEs for six or more components \cite{Axel6C,ChemicalEmbedding,10C_with_embedding}, all involving more well-behaved alloy systems.
Accordingly, the focus of this work is to present an active learning workflow for CE construction that explicitly  incorporates instability avoidance. 
We detail the complete workflow: 
(1) the overall active learning loop, 
(2) ensemble CE model formulation and training, 
(3) active structure selection, filtering criteria, and Mahalanobis distance-based data diversification, 
(4) DFT calculation details, 
and (5) implementation considerations. 
We analyze the training yields, CE model performances, the effectiveness of average atomic displacement as a stability classifier, and data diversification. 
We conclude by assessing the final model quality and potential improvements.

\begin{figure*}[t]
\centering

\includegraphics[width=0.9\linewidth]{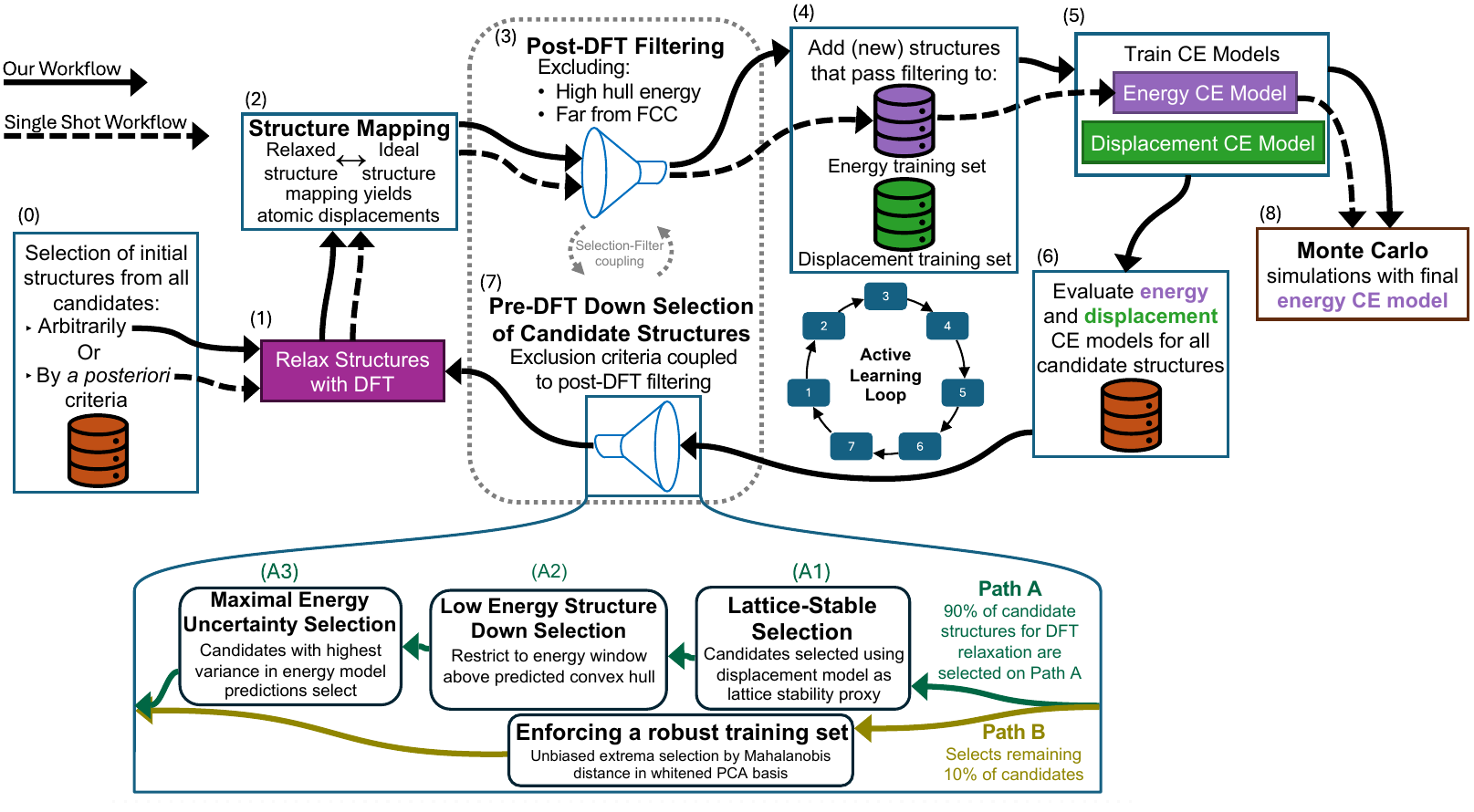}
\caption{\label{fig:figure_active_learning_loop} 
Schematic of the active learning workflow, with initial structure selection as the entry point to the loop. 
The pre-DFT structure selection step, which is designed to maximize throughput by targeting structures likely to pass post-DFT filtering, is highlighted.
(Inset) Overview of the candidate structure selection strategy.
In primary Path A, candidates are first filtered by predicted energy relative to the convex hull, then by predicted FCC stability using the displacement model, before the remaining structures are ranked and selected based on energy uncertainty.
In the secondary Path B, structures are selected solely based on their Mahalanobis distance from the existing training dataset to promote data diversity.}
\end{figure*}

\section{Workflow}

\subsection{Overview}

The distinguishing feature of our  workflow is the explicit pre-DFT targeting of structures likely to survive post-DFT filtering, thereby minimizing wasted DFT calculations. 
To realize this, our active learning loop is organized as into seven steps, illustrated schematically in \autoref{fig:figure_active_learning_loop}.
Entry into the loop begins with Step 0, where an initial set of structures is selected according to predefined criteria. 
In Step 1, the selected structures are relaxed using DFT calculations.
Step 2 maps the relaxed structures back onto decorations of the ideal lattice. This mapping also yields both the discrete latttice decoration and the associated atomic displacements from ideal lattice sites.
In Step 3, the relaxed structures are filtered according to two post-DFT criteria: formation energy relative to the thermodynamic convex hull and average atomic displacement. 
These filters determine whether a structure is admissible for training the energy CE and/or displacement CE models. 
In Step 4, structures that pass the filtering criteria are added to the training data for the energy CE and/or displacement CE models. 
Step 5 consists of training ensemble models for both the energy CE and displacement CE using the updated datasets.
In Step 6, all candidate structures are evaluated using the trained CE models to predict hull energy, mean atomic displacement, and their models'  uncertainties. 
%
To close the active learning loop, Step 7 uses two parallel paths, A and B, to select new structures for DFT evaluation based on predictions from the previous step, the post-DFT filtering criteria, and the current training training dataset distribution. Step 7 is fully detailed in the active structure selection and filtering section below. 
%


\subsection{Model Formulation and Training}

The cluster function vector space is defined by all symmetry-inequivalent clusters permutations on an ideal FCC lattice with lattice parameter $a= 3.6$\,\AA.
For pairs, triplet, quadruplet, and quintuplet clusters, the corresponding cutoff radii used in this work are 8.0\,\AA, 4.5\,\AA, 3.6\,\AA, and 3.6\,\AA, respectively.
These cutoffs correspond to maximum neighbor shells of the 9\textsuperscript{th},  7\textsuperscript{th}, 3\textsuperscript{rd}, and 1\textsuperscript{st} nearest neighbors, respectively.
With these truncation choices, the number of unique cluster permutations for each body order is 150 (pairs), 495 (triplets), 415 (quadruplets), and 600 (quintuplets), respectively. 
Our ensemble cluster model is constructed using  bootstrap aggregating, with the final prediction taken as the average over the ensemble. The predicted property is given by: 
\[ Q= \bar{\textup{J}}^{ T } \Gamma\,,  \]
where $Q$ is the targeted scalar property (mixing energy or average atomic  displacement in this work), $\bar{\textup{J}}$ is a column vector of the averaged cluster interaction parameters (effective cluster interactions, ECIs, multiplied with their cluster multiplicities), and $\Gamma$ is the cluster function vector, also referred to as the cluster correlation vector \cite{compressive_sensing_Hart} or basis function vector \cite{icetTutorial2024}).
The uncertainty in the ensemble prediction is quantified by the standard deviation of model outputs, 
which can be written as
\[ \sigma_{Q}^{2} =  \Gamma^{T} \Sigma_{ \textup{J}} \Gamma \,,\] 
where $\sigma_{Q}$ is the standard deviation of the predicted property and $\Sigma_{ \textup{J}}$ is the symmetric covariance matrix of cluster interaction  parameters vectors across the ensemble, 
$\Sigma_{ \textup{J}} =  \mathrm{Cov}\left(  \textup{J}, \textup{J} \right)$. 
\begin{table*}[t]
    \centering
    \caption{Selection criteria (pre-DFT, using predicted values) and filtering criteria (post-DFT, using DFT values) for the energy and displacement models}
    \begin{tabular}{lcc}
        \toprule
        & Hull Energy, {$E_{\mathrm{hull}}$} (meV/atom) & Displacement, {$\delta$}  (\AA) \\
        \midrule
        Selection criteria (predicted) 
            & $E_{\mathrm{hull}} + \sigma_E < 200$ 
            & $\delta_{\mathrm{avg}} + \sigma_{\delta} < 0.2$ \\
        Filters for energy model (DFT) 
            & $E_{\mathrm{hull}} < 200$ 
            & $\delta_{\mathrm{avg}} < 0.2,\ \delta_{\max} < 0.4$ \\
        Filters for displacement model (DFT) 
            & $E_{\mathrm{hull}} < 400$ 
            & $\delta_{\mathrm{avg}} < 0.4,\ \delta_{\max} < 0.8$ \\
        \bottomrule
    \end{tabular}
    \label{tab:table_selection_and_filtering_criteria}
\end{table*}

\subsection{Active Structure Selection and Filtering}
To enter the active learning loop, an initial dataset is required (Step 0). 
Although \textit{a priori} model-based initialization schemes are possible, we adopt a  general and unbiased approach by selecting all SIS of size four and smaller (1,536 total structures). 
These structures are relaxed using DFT (Step 1). After which, a structure-mapping procedure determines the extent to which each relaxed structure deviates from the ideal FCC lattice (Step 2). 
Structures that relax sufficiently far from the ideal lattice are filtered out (Step 3) prior to inclusion in the training datasets (Step 4).
Structural instability is quantified using two metrics: ($i$) the average atomic displacement,  $\delta_\mathrm{avg}$, and ($ii$) the maximum atomic displacements,  $\delta_\mathrm{max}$, relative to the ideal lattice. 
Contributions from cell rotation, uniform cell volume changes, and net atomic translations are removed before computing these quantities.
Likewise, structures that are energetically unlikely  to be relevant for the temperatures of interest in subsequent MC simulations are also filtered using the energy above the convex hull, $E_\mathrm{hull}$, or hull energy. 
The criteria used for post-DFT filtering and pre-DFT selection for both the energy and displacement models are summarized \autoref{tab:table_selection_and_filtering_criteria}.
Once the initial energy and displacement models are trained (Step 5), the active learning loop proceeds by  evaluating all candidate structures using the current ensemble models (Step 6).
These predictions  inform the pre-DFT down selection of 1,000 structures (per loop) in Step 7, which is carried out via two complementary selection paths, A and B.
Path A comprises structures selected to maximize informational gain while explicitly targeting those likely to pass post-DFT filtering. 
Using the current iteration of ensemble models, candidates predicted to exceed tolerance thresholds (\autoref{tab:table_selection_and_filtering_criteria}) for average atomic displacement (A1) or energy above the predicted convex hull (A2) are excluded. 
Among the remaining candidates, those with the largest predicted energy uncertainty are selected for DFT calculations (A3). 
This first Path A accounts for 90\% of the selected structures in each iteration.
To further reduce the likelihood of discarding structures after DFT relaxations, the selections cutoffs are made conservative by incorporating by predicted ensemble uncertainties. 
Specifically, only candidate structures satisfying a predicted hull energy plus energy uncertainty less than the post-DFT energy filter threshold of 200 meV/atom are retained (see row 1, Table~\ref{tab:table_selection_and_filtering_criteria}).
The degree of conservativeness can be adjusted by scaling the ensemble uncertainty contribution.
After applying both energy and displacement selection criteria, the remaining candidates are ranked by energy uncertainty, and the 900 structures with the highest uncertainty form the uncertainty-selected subset.
The displacement model employs more permissive post-DFT filtering thresholds than the energy model, with cutoff values doubled relative to those used for the energy model.
This choice reflects the role of the displacement model as a stability classifier rather than a high-accuracy predictor.
The post-DFT displacement filters include a maximum displacement cutoff set to $2\times$ the average displacement cutoff. 
While this 2:1 is consistent with trends observed in the maximum:average ratio in our data set, in practice, the maximum displacement criterion excludes few additional structure beyond those filtered by the average average displacement metric alone.
Path B comprises  the remaining 10\% of structures and is focused on improving data diversity. 
Here, whitened principal component analysis (WPCA) is performed on the cluster function vectors of the current training set.
This transformation enables scoring of candidate structures by their unbiased distance to the training set distribution, quantified by the Mahalanobis distance, and allows tracking of the dimensional span of the sampled configurational space. 
The biasing scheme strongly rewards projections along poorly sampled directions in the cluster-function vector space while weakly weighting directions that are already well represented. 
The resulting selection behavior is similar to  structure orthogonalization procedures used in compressed sensing approaches \cite{Bayesian_compressive_sensing_Hart}. 
For Path B,  candidate structures not already selected via Path A are ranked purely by their Mahalanobis distance from the existing energy training set distribution, and the top 100 structures are selected. 
Additional details of the WPCA-based selection procedure are provided in the SI, Section 3.
Beyond enhancing data diversity, the Path B structure selection serves as a safeguard against any potential failure modes of active learning, in which reliance on an evolving model to select new training data can lead to  self-reinforcing biases. 
Conversely, selection schemes driven solely by extremal criteria may increasingly target low-yield structures as the training set expands.
The reduced training data yield expected from the diversity-selected group is therefore considered an acceptable tradeoff for improved robustness, particularly since this path affects only 10\% of the selected structures per iteration.
\subsection{DFT Parameters}
Density functional theory (DFT) calculations were performed with the Vienna Ab initio Simulation Package, \texttt{VASP}, (version 6.2.1) using the Perdew-Burke-Ernzerhof (PBE) exchange-correlation functional \cite{PBE}.
The PBE pseudopotentials used were Ni:02Aug2007, Fe:06Sep2000, Ti\_pv:07Sep2000m Ni:02Aug2007, Fe:06Sep2000, and Ti\_pv:07Sep2000 from the set of pseudopotentials provided with \texttt{VASP}. 
All calculations were initialized in a ferromagnetic configuration, with an initial 2\,$\mu_{B}$ applied to each atom.

Since large cell shape changes were anticipated for a subset of the structures during relaxation, ionic relaxations were intentionally limited to a small number of steps per run to allow for refreshing of the planewave basis and updating of the \textit{k}-point grid. 
$\Gamma$-centered \textit{k}-point grids were employed and selected to maintain a minimum reciprocal space \textit{k}-point density of 20,000 $\text{\AA}^{3}$. 
This value was chosen as a compromise between maintaining energy continuity during cell relaxation-induced \textit{k}-point grid updates and overall computational cost. 

Prior to relaxation, all enumerated structures were perturbed to break ideal lattice symmetry. 
Cell vectors were perturbed by multiplication with a matrix $\mathbf{A} = \mathbf{I} + \mathbf{\epsilon}$, where $\mathbf{I}$ is the identity matrix and the elements of the symmetric matrix, $\mathbf{\epsilon}$ were drawn from a normal distribution with a standard deviation of 0.07.  
After cell perturbation, atomic Cartesian  coordinates were displaced randomly using normally distributed displacements with a standard deviation of 0.2\,\AA. 
The perturbed structures are then relaxed until the maximum force on any atom was less than of 5 meV$\text{\AA}^{-1}$. 
Structure relaxations were aborted after 500 iterations. 
Structures that failed to achieve the maximum atomic force criterion for relaxation convergence were excluded from all datasets. 
Such failure were commonly associated with large cell shape changes that would map poorly onto the target FCC lattice and were therefore unsuitable for the CE training. 
All remaining structure-independent DFT input parameters are provide in the SI (Table S2).
We also note that the interpretation of effective cluster interaction (ECIs) is challenging for such a high-dimensional model.
Accordingly, these plots are provided in Figures S2, S4, and S6, while alternative visualization approaches for complex systems remain an area for future work.
\subsection{Implementation Details}
The workflow is built on the modular framework provided by the \texttt{icet} package \cite{icetPaper2019}, which is used for cluster function vector space construction, structure mapping, and SIS generation via structure enumeration.
The atomic simulation environment (\texttt{ASE}) \cite{ASE} is used for DFT job management, and ASE databases are used to tag, track, and organize training data.
Model training and fitting are carried out using the \texttt{trainstation} package \cite{Trainstation_2020}. 
Ensemble model training (Step 5) was performed using \texttt{trainstation}'s \texttt{EnsembleOptimizer} module. 
Each ensemble contained 100 individual CE models.
All models are fit using LASSO regularization with a minimal regularization parameter of $\alpha=10^{-4}$, which was found necessary to prevent numerical instabilities in under-determined fitting scenarios. 
Lower values of  $\alpha$ resulted in convergence failures during optimization.

For each individual model within the ensemble, bootstrap aggregating is performed by randomly selecting 90\% of the available training dataset for fitting.
The remaining 10\% of structures are assigned to that model's test set.
The ensemble cross-validation root-mean-square error (CV RMSE) is reported as the mean CV RMSE across all individual models in the ensemble.
%

\section{Results}
 
\subsection{Yields}
To characterize the dataset generated by the active learning workflow, 
we decompose the collected structures into a hierarchy of subsets defined by increasingly strict criteria. 
Of the actively selected SIS, those that completed DFT relaxation within the allotted number of ionic optimization steps are classified as \textit{relaxed} structures.
From this set, structures that can be mapped back onto an FCC lattice form the \textit{FCC-mappable} subset.
Within the FCC-mappable set, structures that  both displacement- and hull energy-based filtering criteria listed in bottom row of \autoref{tab:table_selection_and_filtering_criteria} constitute the  \textit{displacement training set}. 
Applying the stricter displacement criteria used for the energy model defines the \textit{FCC-stable} subset. 
Finally, structures that satisfy both the displacement and hull energy filtering criteria for the energy model, middle row of \autoref{tab:table_selection_and_filtering_criteria}, form the \textit{energy training set}.

\begin{table}
    \centering
    \caption{Structure yields at successive filtering stages of the active learning workflow. Yields are reported relative to the total number of  selected structures.}
    \label{tab:loop_data_mean_yields}
    \begin{tabular}{lll}
        \toprule
        Subset & Structures ($N$) & Yield (\%) \\
        \midrule
        Selected                  & 11{,}536 & 100.00 \\
        Relaxed                   & 11{,}403 &  98.85 \\
        FCC-mappable              & 10{,}084 &  87.41 \\
        Displacement training set &  9{,}024 &  78.22 \\
        FCC-stable                &  7{,}708 &  66.82 \\
        Energy training set       &  5{,}309 &  46.02 \\
        \bottomrule
    \end{tabular}
\end{table}

The overall yields for each of these subsets are summarized in  \autoref{tab:loop_data_mean_yields}.
Despite the fact that only two of the six constituent elements are stable in the FCC structure and that numerous non-FCC intermetallic phases are known in this composition space, we observe a surprisingly high mean yield of 66.82\% FCC-stable structures following DFT relaxation.
The average yield of energy training data is lower, at 46.02\% (5,309 structures), reflecting the additional energetic filtering requirements.  

\begin{figure}
\centering
\includegraphics[width=0.92\linewidth]{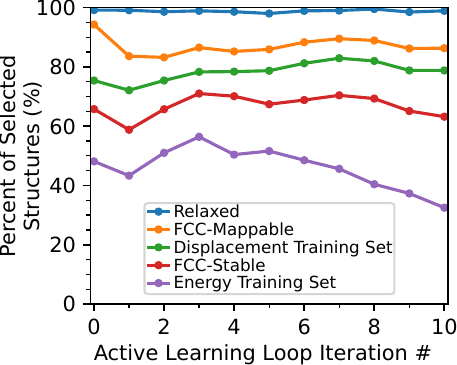}
\caption{\label{fig:figure_yields_vs_loop} Percentage yield of actively selected structures belonging to each dataset subset as a function of active learning iteration.}
\end{figure}

\begin{figure*}[t]
\centering
\includegraphics[width=0.95\linewidth]{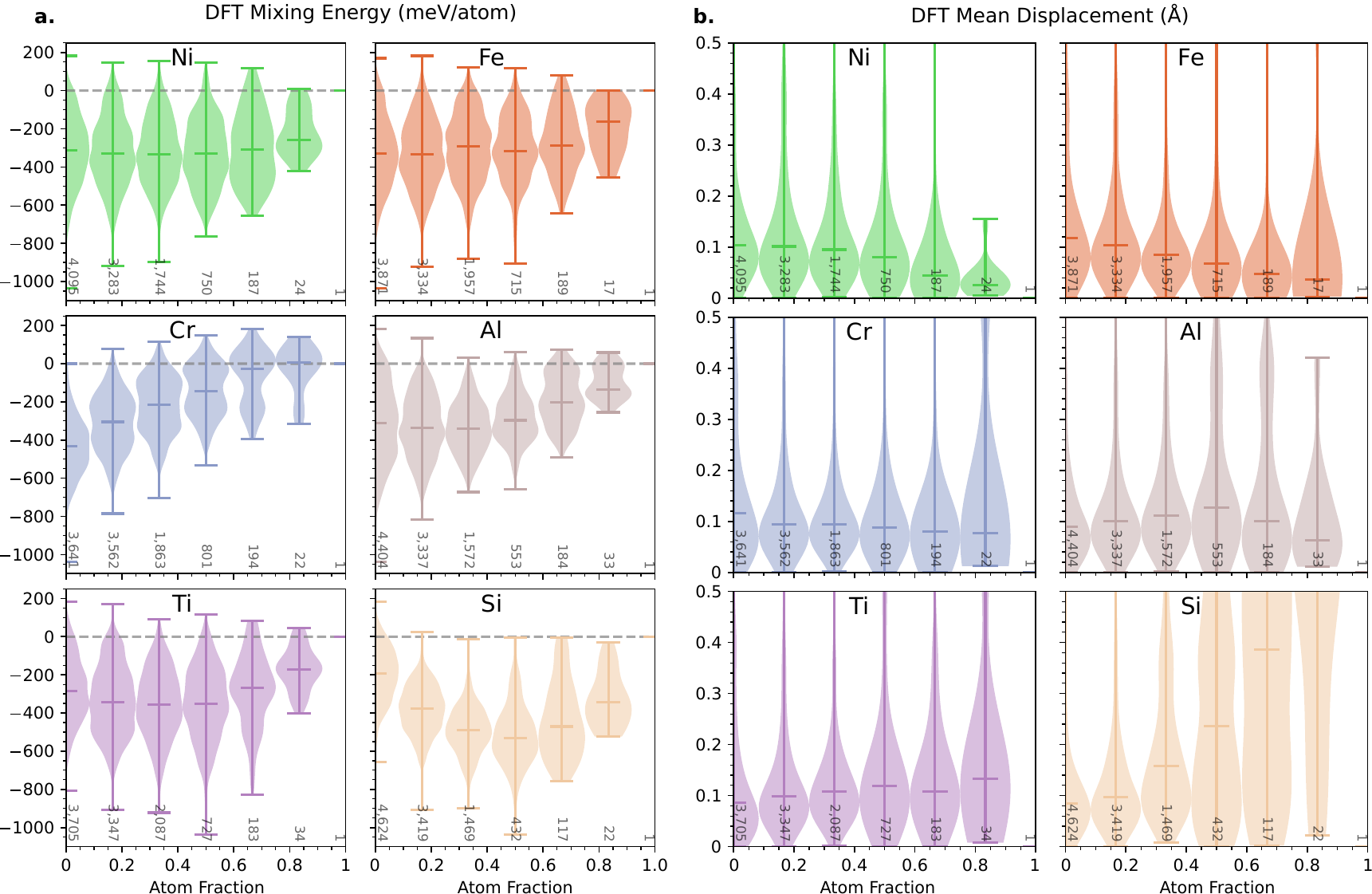}
\caption{\label{fig:figure_loop_data_distributions} 
Distributions of DFT-level properties for FCC-mappable structures resolved by elemental composition.
(a) Mixing energy and (b) mean atomic displacement from ideal FCC, shown as histograms (or violin plots) for bins of atomic fraction of each element.
Horizontal bars indicate the minimum, median, and maximum values within each bin.
Bin edges are defined in intervals of $1/6$, such that the first bin contains structures with zero concentration of the selected element and the final bin contains structures composed entirely of that element.
The total number of structures per bin, which varies by up to three orders of magnitude, is indicated in gray to provide context for the distribution statistics.}
\end{figure*}

To better understand this discrepancy, subset  yields are resolved by active learning loop number (\autoref{fig:figure_yields_vs_loop}).
The initial selection of small SIS (size $\leq$ 4) exhibits particularly high FCC-stable yields, followed by a decline during  the first active learning iteration and a partial recovery in subsequent loops.
We attribute this behavior to the limited ability of smaller supercells to accommodate long-wavelength unstable modes (when present), in contrast to larger supercells that can more readily accommodate such instabilities.
After the initial recovery, all subset yields remain relatively stable with loop number except for the except the energy training set which shows a steady decline in yield.
Because the FCC-stable yield, which already satisfies the average displacement criterion associated with the energy model, remains approximately constant, this decline must be associated primarily with energetic filtering.
We therefore hypothesize that the yield reduction arises from progressive depletion of available SIS configurations that fall within the targeted energy window near the convex hull as the model improves.
To examine elemental correlations underlying these trends, we decompose the DFT-level FCC-mappable subset into atom-fraction-binned  histograms for each element, shown in \autoref{fig:figure_loop_data_distributions}. 
Histogram bins are defined using exclusive lower and inclusive upper bounds, \textit{i.e.}, $x_{\mathrm{low}} < x \leq x_{\mathrm{high}}$.
The first bin in each histogram corresponds to  structures lacking the given element, while subsequent bins track the trend effect of increasing elemental concentration from the most dilute limit (in the set) toward the pure-element limit.
For DFT-level mixing energies (\autoref{fig:figure_loop_data_distributions}a), increasing concentrations of Ni, Fe, and Al have relatively modest effects on the median mixing energies. 
Cr has low mixing energies with the other elements, leading to progressively lower median mixing energies as the Cr fraction increases. 
Ti shows a pronounced minimum in median mixing energy near 1/3 atomic fraction, consistent with its strong tendency to form intermetallic compounds.
Si exhibits an even stronger propensity for silicide formation, with the median mixing energy occurring near 50\% Si. 

The DFT-level mean displacement from ideal FCC is shown in  \autoref{fig:figure_loop_data_distributions}b. 
The most striking feature is the sharp increase in displacement above 20\% Si, clearly highlighting the destabilizing role of Si in the FCC lattice in this alloy space. 
Increasing concentrations of Ni, Fe, and Cr tend to reduce median displacements, while Ti  increases them.  
Al exhibits a peak in displacement near 50\% concentration, which may be associated with a tendency to form B2-type phases with Ni and Fe and a tetragonal analog with Ti.

Consistent with these trends, the total number of FCC-mappable structures with Si concentrations above 20\%  (2,501 structures) is substantially lower than for the same concentration range applied to the other elements: Ni 3,159, Fe 3,332, Cr 3,342, Al 2,798, and Ti 3,470. 
This reduction is due to both increased lattice instabilities, leading to unmappable structures, and the active avoidance of high-displacement configurations during the pre-DFT selection.

\subsection{Performance Progression}

\begin{figure}[t]
\centering
\includegraphics[width=0.88\linewidth]{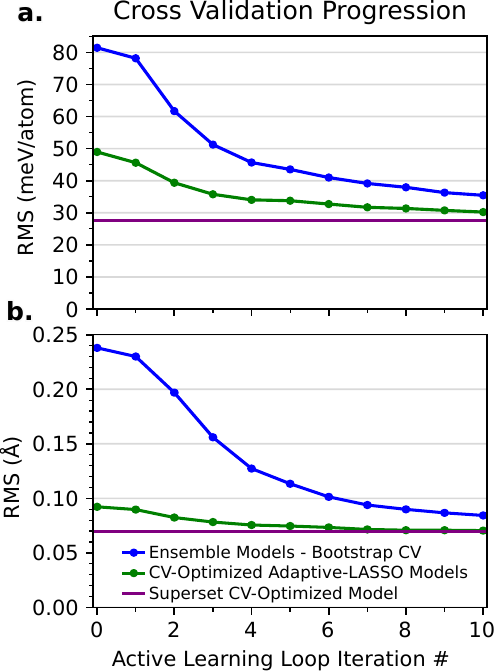}
\caption{\label{fig:figure_performance_vs_loop} Evolution of cross-validation performance as a function of active learning loop iteration for (a) mixing energy models and (b) average atomic displacement models.}
\end{figure}

With each active learning iteration, the addition of newly relaxed structures improves the predictive performance of the ensemble models, as reflected by the steadily decreasing cross-validation (CV) scores in \autoref{fig:figure_performance_vs_loop}. 
In addition to ensemble models, we also fit models using an adaptive LASSO scheme with logarithmically varying levels of $L^{1}$ regularization to find the cross-validation-optimal (CV-optimal) model at each learning loop. 
While the ensemble models provide uncertainty sensitivity through variance across members and implicit regularization through  averaging, applying strong regularization during ensemble fitting  would suppress inter-model variance and thereby degrade overall uncertainty estimates. 
Thus, the CV-optimal regularized models give a soft lower bound on the best achievable CV performance attainable with a given training set. 
This comparison is useful for assessing when model accuracy has reached a level sufficient to terminate the active learning loop.
In principle, a  Bayesian formulation could simultaneously provide both uncertainty estimates and appropriate regularization within the a single model, as demonstrated in Refs.\ \cite{MuellerBayesian,mueller_exact_expressions_for_selection,AxelBayesian}. 
Because the present workflow is agnostic to the underlying model formulation, incorporation of such approaches is left for future work.

Finally, we also report CV-optimal models trained on a superset of data assembled from multiple previous training attempts ($\sim 2 \times$ the final dataset size) as an estimate of the asymptotic lower bound on achievable CV error.
Except for these superset models, all CV scores are computed using only the data available at each learning loop, reflecting the  actionable, in-progress model performance rather than retrospective  analysis.  
For the ensemble energy models, the largest improvements occur during approximately the first five learning loops, after which improvements continue more gradually. 
This behavior is expected for a chemically and configurationally complex system.
In contrast, the ensemble displacement models rapidly improve and then stagnate. The corresponding CV-optimal regularized displacement models start close to the  superset model performance and show only marginal  improvement with additional training data.
The rapid recovery in FCC-stable yields observed in \autoref{fig:figure_yields_vs_loop} is therefore most likely due to the rapid improvement of the displacement model during the early learning loop iterations  (\autoref{fig:figure_performance_vs_loop}b).
This rapid improvement is likely due to two primary factors: (1) the displacement models use more permissive filtering thresholds, resulting in a larger volume of usable training data added per learning loop. 
(2) The average atomic displacement is an intrinsically simpler property to model than mixing energy.
As evidence for the latter, the CV-optimal superset  energy model has 527 non-zero parameters derived from 9,984 training structures, while the corresponding displacement model contains  410 non-zero parameters trained on 17,310 structures. 
The CV-optimal superset energy model has more than double the number of parameters per training structure, highlighting its greater representational complexity and slower convergence behavior.

\subsection{Stability Classified by Displacement}

\begin{figure}
\centering
\includegraphics[width=0.99\linewidth]{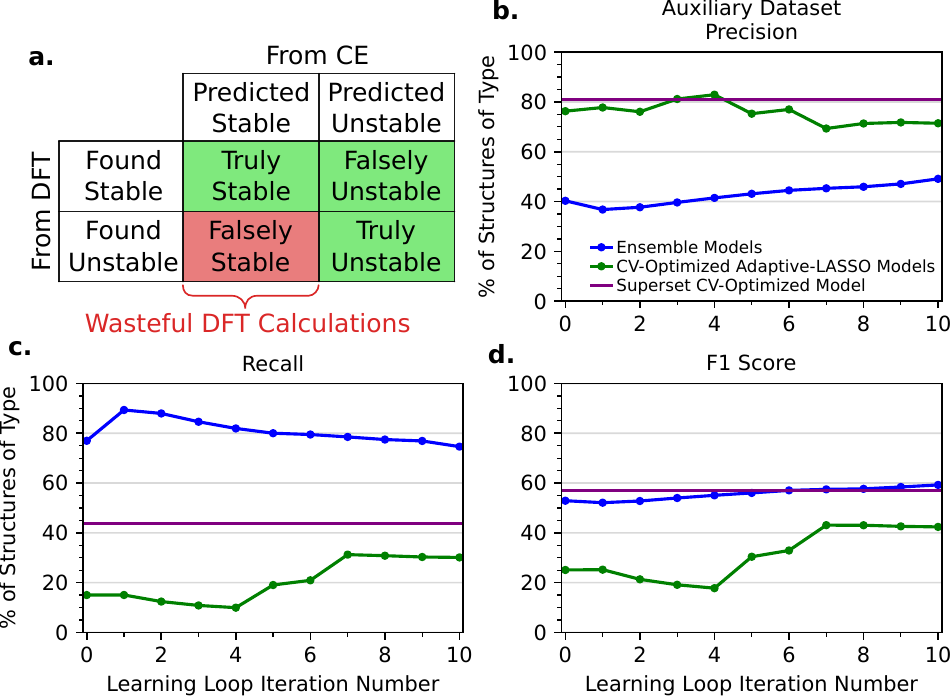}
\caption{\label{fig:stability_classified_by_displacement}
Stability classification performance based on predicted average atomic displacement. (a) Confusion matrix defining stability classification outcomes. (b) Precision, (c) recall, and (d) $F_1$ score as a function of active learning iteration for ensemble-based classifiers, per-loop CV-optimal models, and the superset CV-optimal model. Metrics are evaluated using only the auxiliary dataset not included in model training.}
\end{figure}

The combination of displacement CE models with explicit thresholds is equivalent to constructing a structural stability classifier model.
This classifier excludes structures that are likely to relax far from the FCC lattice prior to DFT evaluation, thereby reducing wasted computational effort.
To quantify classifier performance, we employ common classification metrics: precision, recall, and $F_{1}$ score defined as:
\begin{align*}
\mathrm{Precision} &= \frac{ \mathrm{Truly \; Unstable}}{\mathrm{Falsely \; Unstable} + \mathrm{Truly \; Unstable}  }\\
\mathrm{Recall}    &= \frac{ \mathrm{Truly \; Unstable}}{  \mathrm{Falsely \;   Stable} + \mathrm{Truly \; Unstable} } \\
F_{1} &= 2 \frac{\mathrm{Precision} \cdot \mathrm{Recall}}{\mathrm{Precision} + \mathrm{Recall}} \,,
\end{align*}
where \emph{truly unstable} denotes structures correctly classified as unstable while \emph{falsely stable} denotes structures incorrectly predicted to be stable.
The four possible outcomes of the classification are shown in the confusion matrix in \autoref{fig:stability_classified_by_displacement}a.
Among these outcomes, falsely stable structures are unusable and lead directly to wasteful DFT calculations.
In contrast, falsely unstable structures are less consequential, since they remain in the candidate pool and may be reconsidered in subsequent learning iterations.
Achieving high recall is the primary objective of the stability classifier. 
The ensemble-based displacement classifiers consistently achieve high recall because the classification incorporates a 
tolerance factor proportional to the predicted uncertainty defined in \autoref{tab:table_selection_and_filtering_criteria}.
In contrast, stability classification  based on the 
per-loop CV-optimal models or the superset CV-optimal model performs noticeably worse.
These regularized models do not provide uncertainty estimates, preventing conservative thresholding and underscoring the importance of uncertainty-aware model formulations for pre-DFT filtering.

\subsection{Diversification}
\begin{figure}
\centering
\includegraphics[width=0.88\linewidth]{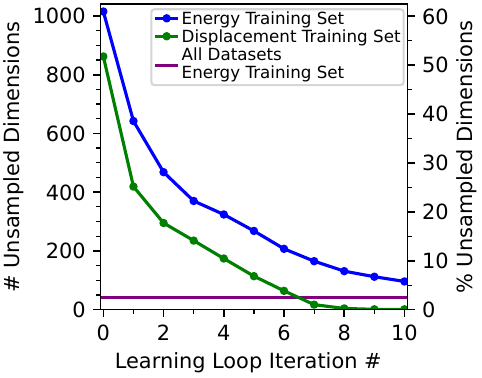}
\caption{\label{fig:figure_diversity_vs_loop} 
Evolution of training data coverage in cluster-function space as a function of active learning iteration.
Shown is the number of unsampled (unspanned) dimensions for the displacement and energy training sets, illustrating the effect of Mahalanobis-distance-based diversification on dataset completeness.}
\end{figure}

As part of the Mahalanobis distance-based data diversification strategy, we explicitly track the dimensional span of the training datasets in cluster-function space across active learning iterations. 
Specifically, we examine the complementary metric: number of unsampled dimensions remaining in the training set as a function of learning loop iteration.
Results for both the displacement training set and the energy training set, which is a subset of the former, are shown in \autoref{fig:figure_diversity_vs_loop}.
The displacement training set rapidly expands its coverage of cluster-function space and achieves full dimensional spanning by the ninth learning loop.
In contrast, the energy training set continues to improve, increasing its coverage throughout the workflow, and reaches a final 95 unsampled dimensions, corresponding to 5.7\% of the total space.
For reference, the CV-optimal superset energy training dataset contains only 41 unsampled dimensions (2.5\%), reflecting the benefit of substantially larger training data volume.
As expected, the superset displacement training dataset inherently spans the full cluster-function space, since the displacement training set itself already achieves complete coverage.
These results demonstrate that the Mahalanobis-based diversification strategy is effective at systematically expanding the sampled configuration space, particularly for the displacement model, and provides steady improvement for the more selectively filtered energy training set.

\subsection{Ultimate Models}
\begin{figure}[t]
\centering
\includegraphics[width=0.96\linewidth]{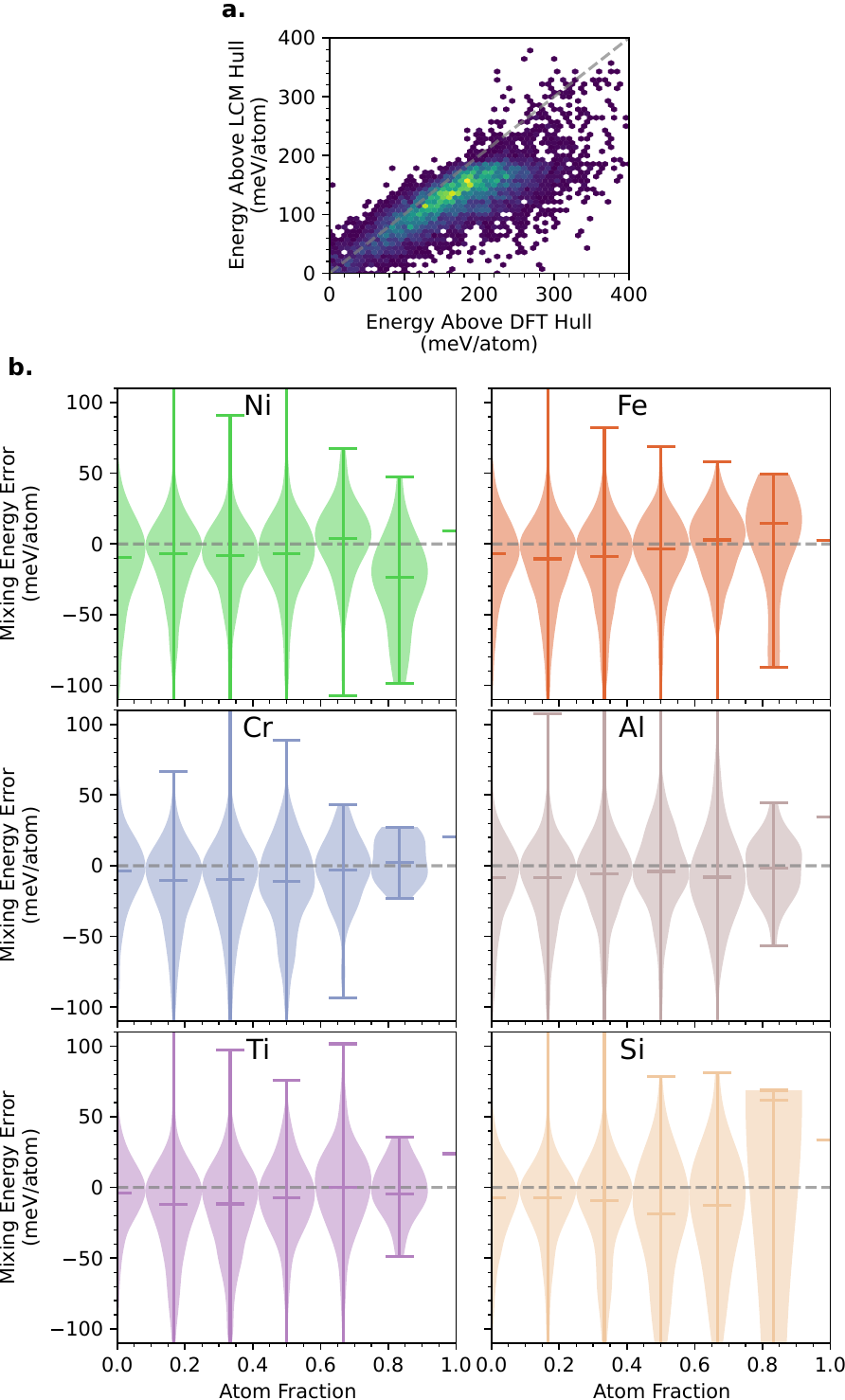}
\caption{\label{fig:figure_final_energy_model} 
(a) Parity plot (shown as a density heat map) comparing predicted and DFT-calculated energies above the convex hull.
(b) Mixing energy prediction errors resolved by elemental composition, shown as histograms binned by atomic fraction; horizontal bars indicate the minimum, median, and maximum values within each bin for the FCC-stable subset.}
\end{figure}

The final ensemble energy model achieves a cross validation (CV RMSE score of 35.4\,meV/atom. 
The parity plot of predicted versus DFT-calculated hull energies shows that the model systematically under-predicts energy above the convex hull for structures more than $\sim$100\,meV/atom above the hull (\autoref{fig:figure_final_energy_model}a). 
Thus, the model slightly overestimates thermodynamic stability in the high-energy regime. 
The element-resolved distributions of mixing energy errors shows strong correlations with composition (\autoref{fig:figure_final_energy_model}b). 
In particular, errors increase significantly with increasing Si content, reflecting the difficulty of accurately modeling the strong compound-forming behavior of Si.
At high concentrations of Ni and Fe, the mixing energy error also increase.
While compositional regions dominated by a single element are inherently under-sampled due to combinatorial constraints, this effect alone cannot explain the observed trends, as similar behavior is not seen for weakly magnetic elements.
This suggests that magnetic interactions in Ni-  and Fe-rich compositions as a likely contributing factor.
In simpler compositions spaces, it is common to compare predicted and DFT binary convex hulls, 
Given the 15 binary systems present in this work, however, these comparisons are presented in Figure S3 for brevity. 
A consistent trend is that the relative errors are larger for binaries with small mixing energy scales like Ti-Cr and smaller for those  with high mixing energy binaries like Ti-Al. 
This behavior is expected in binaries like Ti-Cr because the intrinsic magnitude of mixing energies is comparable to the model CV error, and because compositional edges (binary limits of the 6-component simplex) are comparatively under-sampled. 
Excluding high-Si configurations from the training set is unlikely to significantly improve model performance.
The RMSE prediction error for all training structures is 22.5 meV/atom compared to 21.9 meV/atom when restricted to structures with $\leq$ 20\% Si, indicating that Si-related error contributions are not dominant in the overall CV metric.

\begin{figure}[t]
\centering
\includegraphics[width=0.96\linewidth]{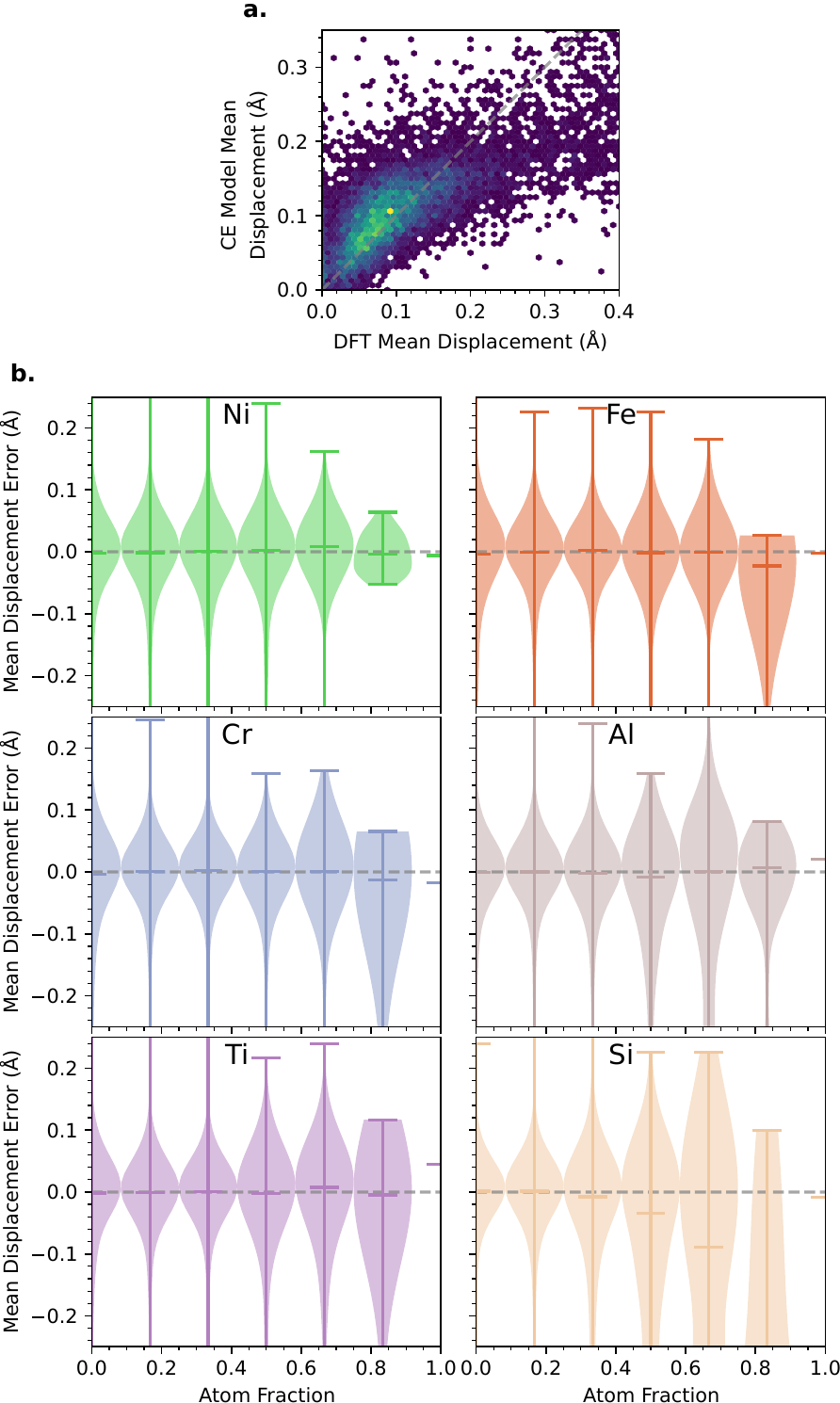}
\caption{\label{fig:figure_final_dravg_model} 
Performance of the final ensemble displacement model.
(a) Parity plot comparing predicted and DFT-calculated average atomic displacements from the ideal FCC lattice.
(b) Prediction errors in average displacement resolved by elemental composition, shown as histograms binned by atomic fraction with horizontal bars indicating minimum, median, and maximum  for the FCC-mappable subset.}
\end{figure}

The final ensemble displacement model achiveve a CV RMSE of 0.084\,\AA. 
The parity plot in \autoref{fig:figure_final_dravg_model}a shows a slight systematic over-prediction of average atomic displacement. 
On a relative scale, the displacement model exhibits lower predictive accuracy than the energy model despite having a larger training set. 
This behavior is expected because average atomic displacement is a derived scalar quantity representing underlying vector displacements and does not correspond to a conventional intensive thermodynamic property.
Consequently, it is less naturally suited to representation within a cluster expansion framework.
Nonetheless, the displacement model provides sufficient predictive fidelity to effectively identify unstable structures  and reduce the number wasteful DFT calculations. 
Furthermore, it may serve as a useful indicator of regions in the Monte Carlo simulations where the assumed FCC lattice becomes unstable.

To estimate asymptotic model performance, we also trained CV-optimal models using a superset dataset combining active learning data with auxiliary data from prior workflow development.
Although these models are not direct products of the present workflow, their larger training sets allow approximation of achievable accuracy.
The distributions of the superset data are similar to those obtained through active learning. 
As the highest accuracy models in this study, the superset models are used for future materials simulations.
The superset energy model was trained on 9,984 structures and achieves an optimal CV RMSE of 27.7 meV/atom using adaptive-LASSO regularization with $\alpha=5\times10^{-5}$ is used. 
Regularization reduces the model to 527 nonzero parameters.
Performance metrics and binary convex hulls analogous to those in \autoref{fig:figure_final_energy_model} and Figure S3 are provided in Figures S4 and S5.
The superset energy model exhibits similar behavior, with a mild tendency to under-predict hull energies (\textit{i.e.}, overestimate stability) as seen in Figure S4a.
The elemental mixing energy error distributions are  largely consistent with those of the ensemble model, with noticeable differences only in high-composition bins containing fewer than 100 structures (Figure S4b).
The superset displacement model, trained on 17,310 structures, achieves a CV RMSE of 0.069 \AA\ 
with adaptive-LASSO $L^{1}$ regularization ($\alpha=5\times 10^{-4}$).
Regularization yields a model with 410 non-zero parameters.
As shown in Figure S6, the superset displacement model displays similar trends to the ensemble displacement model, including a tendency to over-predict displacements and nearly identical element-resolved error distributions.
Assessing the significance of the achieved CV errors of 35.4 meV/atom (final ensemble model) and 27.7 meV/atom (superset model) is nontrivial.
While such values would be considered large for cluster expansions in simpler alloy systems, it remains unclear what level of accuracy is ultimately achievable in a six-component space that includes multiple regions of structural instability.
Direct comparison is further complicated by the limited number of prior studies addressing cluster expansions with six or more components. 
At the time of writing, we are aware of only three such studies.
In the first, Nataraj \textit{et al.}~\cite{Axel6C} developed a cluster expansion for the AlHfNbTaTiZr BCC system, although a CV score was not reported.
However, their cluster expansion for the 5 element subset without Al achieved a CV of 7.8 meV/atom.
Importantly, in this compositional space Nb and Ta are stable in BCC, while Ti, Zr, and Hf can be thermally stabilized in BCC, indicating a relatively well-behaved lattice.
In the second study, M\"uller and Natarajan~\cite{ChemicalEmbedding} constructed both linear and neural network cluster models for the VNbTaCrMoW BCC system using 2,874 structures in their dataset. They achieve a CV ~5.9 meV/atom with 246 ECIs in their linear model, although notably all six elements are stable as BCC. 
In the third study, Lee \textit{et al.}~\cite{10C_with_embedding} extended the work of M\"uller and Natarajan \cite{ChemicalEmbedding} to a 10-component system (TiZrHfVNbTaCrMoW with vacancies) using a chemical embedding neural-network cluster expansion. 
With a dataset of 9,419 structures, they achieved a CV of $~$7.5 meV/atom.
The latter two works highlight the effectiveness of chemical embedding approaches in  reducing the dimensionality of the ECI space for large composition spaces.
In contrast to these studies, the considerably higher CV errors observed in the present work are most plausibly attributed to the presence of extensive regions of structural instability in the Fe-Ni-Cr-Al-Ti-Si composition space.
Near the boundaries of lattice stability, atomic relaxations involve moderate but non-negligible displacements that are only partially captured by a cluster expansion defined on an ideal lattice.
This leads to a fundamental limitation that the cluster basis remains only indirectly sensitive to the configurational energetics associated with emerging instability. 
This behavior can be understood as lying between two limiting cases:
In the first limit, the mapping between ideal lattice decorations and relaxed structures is poor, resulting in weak  correlations between cluster functions and equilibrium energies. 
In this case, the CE performs poorly across the entire alloy space. 
In the second limit, relaxed structures remain close to the ideal lattice decorations, and the CE provides an accurate and efficient representation of the energetics. 
The current alloy system lies in the intermediate regime, where partial lattice instability degrades, but does not fully invalidate, the CE representation. 
The imperfect predictive capability of the displacement model further supports this interpretation.
The lack of identifying directly comparable studies highlights the both the difficulty of the present problem and the need for methodologies, such as the one developed here, that explicitly accounts for instability during model construction.
\subsection{Alternative Active Learning Workflows}
Our use of ensemble models and candidate structure generation via enumeration was  motivated by simplicity of implementation and low per-structure computational cost. 
Following candidate down selection, structures are ranked by their predicted energy variance. 
However, Van de Walle and Ceder suggest that the expected variance reduction is a better metric \cite{atat_automation}, as it directly targets improvement in model accuracy rather than uncertainty alone. 
Bayesian formulations of both variance and expected variance reduction have been developed by Mueller and Ceder \cite{MuellerBayesian,mueller_exact_expressions_for_selection}, as well as an alternative formulation by Chen \textit{et al.} \cite{AxelBayesian}. 
These approaches provide formal, closed-form statistical measures for structure selection and naturally balance exploration and exploitation.
Given these advantages, we suggest that future implementations adopt Bayesian selection strategies.
Related developments in Bayesian compressive sensing \cite{compressive_sensing_Hart,Bayesian_compressive_sensing_Hart} and cluster decomposition methods \cite{ANOVA_Barroso-Luque_Cedar_2024} further point to promising directions for improving model efficiency in high-dimensional composition spaces.
Regardless of the model and selection metric, the generation of candidate structures remains a critical component of any active learning workflow. 
In addition to enumeration, random structure generation combined with duplicate filtering provides a simple alternative for building candidate pools.
More targeted approaches include simulated annealing of ``probe'' structures optimize specific objectives, such as minimizing mixing energy \cite{CLEASE}, maximizing variance reduction \cite{CLEASE}, or targeting specific regions in cluster-function space  \cite{CLEASE,icetTutorial2024}.
A particularly promising direction is the formulation of multi-objective candidate selection strategies.
Such approaches could simultaneously prioritize structures that improve model accuracy, lie near the convex hull, exhibit low predicted displacement, and  expand the sampled cluster-function space.
Developing efficient algorithms to balance these competing objectives represents an important opportunity for future work in active learning for complex alloy systems.

\subsection{Improvements}
Analysis of the ensemble model improvement rates and the number of non-zero parameters relative to training set size in the CV-optimal superset models indicate that displacement models are significantly less complex than energy models.
This suggests that, when used solely as a tool for sampling efficiency, displacement (or instability) models could be simplified by reducing the number of included clusters.
Such reduced-complexity models would converge more rapidly and could improve early-stage active learning by more effectively avoiding unstable structures.
More generally, other scalar properties such as the norm of the lattice strain tensor, which maps  equilibrium cell vectors back to the ideal lattice cell vectors, could be treated analogously to energy within existing CE frameworks. 
Higher-fidelity representations are also possible, including direct expansion of lattice strain tensors or atomic displacement vectors, which could augment or replace the present scalar displacement models for detecting the onset of instability.
For lattice strain tensors, each expanding tensor element as scalar will yield symmetry violating tensors with redundant parameters \cite{tensorial_cluster_expansion_vandeWalle2008}. Thus, symmetry adapted forms of lattice strain tensor cluster expansion are preferable in selecting an existing formulation+implementation. 
The Alloy Theoretic Automated Toolkit (\texttt{ATAT}) is capable of fitting symmetry-adapted cluster expansions for bulk tensorial properties with the theoretical approach developed in Ref.~\cite{tensorial_cluster_expansion_vandeWalle2008}.
Garrity further developed specific formulations for strain-cluster and strain-displacement coupling models \cite{GarrityFC}. 
The most ready to incorporate implementation seems to be the configuration plus strain (CPS) cluster expansion formalism in \texttt{CASM} \cite{CPS1,CPS2,CPS3,CASM}, which provides an efficient route for constructing symmetry-adapted expansions in strain order parameter subspaces.
Extending beyond scalar quantities, improving predictive fidelity by explicitly modeling equilibrium atomic displacement vectors is a natural next step.
Implementations of symmetry-adapted cluster expansion of atomic vector quantities are not common and usually involve the expansion of the potential energy surface in terms of configurations and atomic displacements, \textit{i.e.}, using atomic displacement vectors as inputs rather than predicted quantities.
Examples include  \texttt{hiPhive} \cite{hiPhive}, which expands force constants, and \texttt{spring cluster} \cite{GarrityFC},  which incorporates expansions of force constants, chemical configuration, and magnetism, and \texttt{CASM} \cite{CASM,CASM_anharmonics}, which can expand the potential energy surface in chemical configuration and atomic displacements. 
Within these model formulations, equilibrium atomic displacements can found by minimizing the potential energy by local geometric optimization.
A potentially more direct approach is suggested by the collective cluster deformation (CCD) formalism introduced in Ref.~\cite{CASM_anharmonics}.
In this framework, the potential energy is expanded as sum of the products of the mode amplitudes of the CCD (named collective deformation amplitudes [CDA]) and their respective energy coefficients, 
The mode energy coefficients are cluster expanded in chemical configuration. The associated ECIs for each mode energy coefficient are the model parameters determined by fitting.
To adapt this framework for constructing rigorous models for predicting atomic displacement vectors, it seems the straightforward forward route is to cluster expand the CDA themselves in chemical configuration.
Such an approach could enable direct prediction of displacement fields while retaining symmetry consistency.

Magnetism represents another important avenue for improvement.
Although all DFT calculations in this work were initialized as ferromagnetic, the resulting projected magnetic moments in Figure S1 indicate that most configurations relax to weakly or non-magnetic states, with Fe exhibiting a broad distribution spanning multiple spin states and magnetic orders. 
Incorporating magnetic degrees of freedom directly into the cluster expansion Hamiltonian was avoided here to limit workflow complexity.
However, the inclusion of magnetic information in the dataset provides an opportunity for future development of magnetic cluster expansions.

Finally, the generality of our workflow means that other cluster expansion model formalisms can easily be substituted so long as they also produce reliable uncertainty estimates. 
For this, we suggest one of the Bayesian approaches, 
as they naturally combine uncertainty quantification and regularization.
These methods could be further enhanced by incorporating chemical embedding \cite{ChemicalEmbedding} and by integrating candidate structure generation strategies such as simulated annealing of probe structures \cite{CLEASE} to directly optimize model improvement.

\section*{Conclusions}
We have developed a general active learning workflow for cluster expansion models that autonomously targets low-energy and structurally stable configurations.
This is achieved by augmenting a conventional energy-based cluster expansion with an auxiliary model of average atomic displacement, which serves as an effective proxy for lattice stability during pre-DFT structure selection.
We demonstrated our workflow in the FCC Fe-Ni-Cr-Al-Ti-Si composition space, which presents substantial challenges due to extensive regions of dynamical instability. 
Despite these challenges, and without relying on any  \textit{a priori} composition space restrictions, the workflow achieves a high average yield of 66.82\% FCC-stable structures following DFT relaxation.
The resulting six-component cluster expansion represents, to our knowledge, one of the largest composition spaces explored with a linear cluster expansion model without chemical embedding.
Furthermore, the dataset of 22,938 DFT-relaxed structures constitutes one of the largest reported for cluster expansion studies.
We have made the dataset publicly available to support further development and benchmarking of active learning and cluster expansion methodologies.
Finally, we identified several avenues for future improvement.
These include incorporating additional stability descriptors beyond average atomic displacement, leveraging recent advances in cluster expansion formulations (\textit{e.g.}, tensorial or symmetry-adapted approaches), and adopting Bayesian model frameworks that provide unified uncertainty quantification and regularization.
Together, these directions offer promising pathways toward more robust and efficient modeling of complex, multi-component alloy systems.

\section*{Author contributions}
Michael J. Waters: conceptualization, methodology, software, data curation, formal analysis, investigation, visualization, and writing --- original draft. James M. Rondinelli: writing --- review and editing, funding acquisition, project administration, and supervision.

\section*{Conflicts of interest}
There are no conflicts to declare.

\section*{Data availability}
The datasets supporting this study are openly available at Zenodo (DOI: \href{https://doi.org/10.5281/zenodo.20302412}{10.5281/zenodo.20302412}).
Three datasets are provided.
The first contains the full set of 47{,}720 symmetrically inequivalent supercells (SIS) enumerated for the six-element system with up to six atoms per cell.
The second contains a subset of 22{,}938 structures from the enumerated set that have been relaxed using DFT, along with their corresponding total energies, mixing energies, forces, and atom-projected magnetic moments.
The third dataset consists of the ideal mappings of 
19{,}806 structures to the FCC lattice, including mixing energies and the mapped displacements from ideal.
These mapped structures correspond to the subset of DFT-relaxed structures where mapping was possible.

\section*{Acknowledgments}
The authors gratefully acknowledge funding from the Office of Naval Research through the Multidisciplinary University Research Initiative (MURI) program under award number N00014-20-1-2368.
This research used resources of the National Energy Research Scientific Computing Center, a DOE Office of Science User Facility supported by the Office of Science of the U.S.\ Department of Energy under Contract No.\ DE-AC02-05CH11231 using NERSC award BES-ERCAP0028012.
This research was supported in part through the computational resources and staff contributions provided for the Quest high performance computing facility at Northwestern University which is jointly supported by the Office of the Provost, the Office for Research, and Northwestern University Information Technology. 
We thank Paul Erhart and his group for helpful discussions and for their assistance with the \texttt{icet} software, which forms the foundation of the workflow developed in this work.



\balance


\bibliography{references.bib} 
\bibliographystyle{rsc} 
\end{document}


\maketitle
{\hypersetup{linkcolor=black}\tableofcontents}

\clearpage

\section{Interaction Permutations and Multi-Dimensional Composition Grids}

The formulas for (1) the number of unique permutations of a k-body interaction with $N_{el}$-elements and (2) the total number of composition points in a uniform composition grid spanning a full $N_{el}-1$ dimensional composition space are both related to figurate numbers. Figurate numbers are defined as:
\[ P_d(a) = \binom{a+d-1}{d} =\frac{a(a+1)(a+2)\cdots(a+d-1)}{d!}  \]
where $d$ is the dimensionality, and $a$ is the number of edge points.

The first connection can be highlighted by counting the number of unique pairwise, three-way, and four-way interactions for $N_{elements}$ which are Nth triangular, tetrahedral, and pentatope numbers, respectively. 
Figurate numbers are $k$-body generalization of these $r$-simplex numbers that count unique permutations in which $d=k$ and $a=N_{el}$.
If parameterizing interactions, the complexity scales with the highest $k$-body order as $O \left( N_{el}^{k} \right)$ 

The second connection to figurate numbers is arises when using a regular grid to sample a composition space with $N_{elements}$. 
The dimensionality of the composition space is $d\rightarrow N_{el}-1$, \textit{e.g.} ternary diagrams are 2D triangles. 
With $n_{edge}$ composition points along the grid edges, $a \rightarrow n_{edge}$. For example, $n_{edge}=4$ corresponds to composition points of with elemental fractions of $\left( 0\%, \, 33.3\%, \, 66.7\%, \, 100\% \right)$.
The total number of composition points $N_{comp}$ simplifies to:

\[ N_{comp} \left ( N_{el} , n_{edge} \right ) = P_{N_{el}-1} \left( n_{edge} \right)  =  \frac{\left [  \left ( N_{el} - 1 \right ) + \left ( n_{edge}-1 \right ) \right ]!}{ \left ( N_{el}-1 \right )! \left ( n_{edge}-1 \right )! } \]

Thus, the scaling of the number of composition points is the symmetric for $N_{el}$ and $n_{edge}$, where they individually scale as the power of the other.

\newpage

\section{Enumeration of FCC cells}

Unique enumerations of the FCC primitive cell given in 
\autoref{tab:fcc_permutations} were computed using the structure enumerator available in \texttt{icet} \cite{icetPaper2019}. 

\begin{table}[h]
\caption{\label{tab:fcc_permutations}The number of unique permutations for different numbers of components and supercell sizes for the FCC primitive cell. The BCC primitive cell has same number of unique permutations for each supercell size and number of components.}
\small
\centering
\begin{tabular}{| m{4em}|r|r|r|r|r|r|r|r|r|}
\hline
        & \multicolumn{9}{c|}{No. of Components} \\
\hline 
Supercell Size & 2   & 3     & 4         & 5         & 6       & 7       & 8       & 9       & 10     \\ 
\hline 
        1  & 2      & 3       & 4         & 5         & 6       & 7       & 8       & 9       & 10     \\
        2  & 2      & 6       & 12        & 20        & 30      & 42      & 56      & 72      & 90     \\
        3  & 6      & 21      & 48        & 90        & 150     & 231     & 336     & 468     & 630    \\
        4  & 19     & 96      & 289       & 675       & 1,350   & 2,429   & 4,046   & 6,354   & 9,525  \\
        5  & 28     & 165     & 600       & 1,684     & 3,984   & 8,337   & 15,904  & 28,224  & 47,268 \\
        6  & 80     & 790     & 4,040     & 14,600    & 42,200  & 104,230 & 229,040 & 459,840 & -      \\
        7  & 104    & 1,245   & 8,232     & 37,350    & 130,800 & 380,259 & -       & -       & -      \\
        8  & 390    & 7,482   & 65,960    & 370,795   & -       & -       & -       & -       & -      \\
        9  & 504    & 13,991  & 171,660   & 1,246,350 & -       & -       & -       & -       & -      \\
        10 & 1,211  & 55,461  & 938,454   & -         & -       & -       & -       & -       & -      \\
        11 & 1,364  & 92,565  & 2,119,656 & -         & -       & -       & -       & -       & -      \\
        12 & 7,140  & 734,567 & -         & -         & -       & -       & -       & -       & -      \\
        13 & 5,248  & 863,493 & -         & -         & -       & -       & -       & -       & -      \\
        14 & 18,270 & -       & -         & -         & -       & -       & -       & -       & -      \\
        15 & 33,168 & -       & -         & -         & -       & -       & -       & -       & -      \\
        16 & 95,837 & -       & -         & -         & -       & -       & -       & -       & -      \\
        \hline
\end{tabular}
\end{table}

\newpage

\section{Diversification by WPCA}
Our whitened principal component analysis (WPCA) starts with a standard principal component analysis (PCA) in the cluster function vector space.
The first step is the covariance matrix of the cluster function vectors in the training dataset: 

\[ \Sigma_{ \Gamma} =  \text{Cov}\left(  \Gamma, \Gamma \right) \]

We compute the principal axes (eigenbasis) and principal variances (eigenvalues) of the cluster function vector covariance matrix using the more numerically stable singular value decomposition (SVD): 
\[ \textup{U} \,  \textup{S}  \, \textup{V}^{\star} = \Sigma_{ \mathbf{\Gamma}} \]
where $\textup{S}$ is the diagonal matrix of the principal variances and $\textup{V}^{\star}$ is the eigenbasis. This works with all covariance matrices since they are inherently positive semidefinite. 
We then construct a diagonal whitening matrix to normalize each dimension in the PCA basis:
\[ \textup{W}= \left ( \textup{S}^{\frac{1}{2}}+\eta \, \textup{I}   \right ) ^{-1} \]
where $\textup{I}$ is the identity matrix and a small value, $\eta=10^{-6}$ is added to prevent division by zero where the principal variance values are zero. 
%
%
%
The final transformation from the cluster function vector basis to the WPCA basis is:
\[ \textup{Z} =\left (  \Gamma - \bar{\Gamma } \right ) \, \textup{V} \, \textup{W} \]
where $\bar{\Gamma }$ is mean cluster function vector that points to the center of the distribution and Z is the cluster function vector in the WPCA basis. 
Since any principle direction which is not spanned by the data set will have a principle variance $s=0$ to within rounding error, 
the number of zeros in $\textup{S}$ counts the number of unspanned dimensions in the data set.
The \mhb distance, $\left| \textup{Z}  \right|$, becomes very large, $\sim \eta^{-1}$, when a cluster space vector lies outside the subspace spanned by the current dataset.
This is desirable since we wish to give much more emphasis to new data that will expand the dimensionality of the training data set rather expanding the spread of data in already spanned dimensions.

Our formulation is related to formulations that use the sensing matrix (also known as design matrix \cite{icetTutorial2024},  correlation matrix, or feature matrix \cite{ionic_ce}) of cluster functions, $X$, which is an $N\times m$ matrix of N structures with row vectors containing $m$ cluster functions, with the following formula:

\[ \Sigma_{ \Gamma}+\bar{\Gamma } \bar{\Gamma }^{T} = \frac{ X^{T}X}{N} \]

\newpage

\section{DFT Calculation Settings}

The structure-independent DFT parameters used for running structure relaxations with \texttt{VASP} are shown in \autoref{tab:vasp_parameters}. 


\begin{table}[h]
    \caption{\texttt{VASP} parameters. More cautious values, indicated by $^{\star}$, are used for first relaxation run. }
    \centering
    \begin{tabular}{|c|c| m{20em} |}
    \hline
        Parameter & Value & Comment \\
        \hline
        \hline
  AMIN & 0.0001 & \\
  \hline
 AMIX & 0.1 & \\
   \hline
 BMIX & 0.000001 & \\
   \hline
 ENCUT & 600 & \\
   \hline
 ENAUG & 2400 &  Fine grid 2x finer than coarse grid\\
   \hline
 POTIM & 0.5, 0.2$^{\star}$ &  Relaxation step scale\\
   \hline
 SIGMA & 0.050 & Smearing width\\
   \hline
 EDIFF & 1.00e-09 & \\
  \hline
  EDIFFG & -5.00e-03 & \\
  \hline
 SYMPREC & 1.00e-08 & Ensures no point group symmetries will be used\\
   \hline
 ALGO & Normal & Blocked-Davidson eigensolver\\
  \hline
 GGA & PE & PBE functional\\
  \hline
 PREC & Accurate & \\
   \hline
 IBRION & 2 & Conjugate gradient relaxation\\
  \hline
 ICHARG & 1 & \\
  \hline
 ISIF & 3 & Cell relaxation enabled\\
  \hline
 ISMEAR & -1 & Fermi smearing\\
   \hline
 ISPIN & 2 & Collinear magnetism enabled\\
  \hline
 ISYM & 0 & Only TR symmetry enabled\\
  \hline
 LMAXMIX & 6 & \\
  \hline
 MAXMIX & -30 & \\
  \hline
 NELM & 200 & \\
   \hline
 NELMDL & -6 & \\
  \hline
 NELMIN & 2 & \\
   \hline
 NSW & 12, 8$^{\star}$ & Maximum ionic steps between restarts to ensure planewave basis and k-grid match cutoff and target k-point density\\
  \hline
 NRMM & 4 & \\
   \hline
 LASPH & .TRUE. & \\
  \hline
 LREAL & Auto & \\
\hline
    \end{tabular}
    \label{tab:vasp_parameters}
\end{table}

\newpage

\section{Magnetic Moment Distributions}

All structures were initialized as ferromagnetic spin-up and most finished relaxing in ferromagnetic configurations. However, a small fraction of structures displayed anti-ferromagnetic or ferrimagnetic ordering. The projected magnetic moments were computed by projecting the magnetization density on the Wigner-Seitz spheres around the atomic cores. 
The default Wigner-Seitz radii provided by the \texttt{VASP} pseudopotentials were used. These radii, in Bohr, are 2.430, 2.650, 2.460, 2.500, 2.500, and 2.480 for Ni, Al, Fe, Ti, Cr, and Si, respectively.
The projected magnetic moment distributions for each element across the superset of all structures is shown in \autoref{fig:figure_magmom_distributions}.
%


\begin{figure}[h]
\centering
\includegraphics[width=0.45\linewidth]{si_fig1_magmom_distributions/si_fig1_magmom_distributions_v0.pdf}
\caption{\label{fig:figure_magmom_distributions}  Distribution of projected magnetic moments by element in the superset of all relaxed structures.}
\end{figure}

\newpage

\section{Ensemble Model ECIs}

\begin{figure}[h]
\centering
\includegraphics[width=0.8\linewidth]{si_fig2_final_ensemble_eci_plots/si_fig2_final_ensemble_eci_plots_v1.pdf}
\caption{\label{fig:figure_final_ensemble_eci_plots}  The final ECIs for the (a) energy and (b) displacement ensemble models plotted vs.\ cluster radius.}
\end{figure}

\newpage

\section{Ensemble Model Binary Convex Hulls}
Binary convex hulls are traditionally shown for cluster expansion works, but become cumbersome when there are 15 binary pairs of elements. Binary structures are a small fraction of the data sets since there far fewer structure permutations. For our set of FCC SISs of orbit size 6 and less, there are a maximum of 137 unique binary structures on each edge of the 5D simplex that contains 47,720 unique structures.

\begin{figure}
\centering
Final Ensemble Model
\includegraphics[width=0.9\linewidth]{si_fig3_ensemble_only_fcc_binary_hulls/si_fig3_ensemble_only_fcc_binary_hulls_v0.pdf}
\caption{\label{fig:figure_ensemble_only_fcc_binary_hulls} Binary mixing energies and convex hulls for the FCC-stable subset of structures in the loop dataset at the DFT level and predicted using the final energy ensemble model.}
\end{figure}

\newpage

\section{Superset CV-Optimal Models}
Here we present the CV-optimal model trained on the superset of the active learning loop structures and the auxiliary ones from previous attempts. The performance plots shown in \autoref{fig:figure_superset_energy_model} and \autoref{fig:figure_superset_dravg_model} mirror those in the main text.

\begin{figure}[h]
\centering
\includegraphics[width=0.55\linewidth]{si_fig4_CV-optimal_superset_energy_model/si_fig4_CV-optimal_energy_model_v1.pdf}
\caption{\label{fig:figure_superset_energy_model}Performance metric plots of the superset energy model evaluated on the superset of FCC-stable structures. (a) Predicted energy above hull vs.\ DFT energy above hull, (b) mixing energy error in histograms binned by elemental fraction of each element, (c) energy ECIs plotted vs.\ their respective cluster radii.}
\end{figure}

\begin{figure}
\centering
Superset CV-Optimal Model 

\includegraphics[width=0.81\linewidth]{si_fig5_superset_cvopt_only_fcc_binary_hulls/si_fig5_superset_cvopt_only_fcc_binary_hulls_v0.pdf}
\caption{\label{fig:figure_superset_only_fcc_binary_hulls}  Binary mixing energies and convex hulls for the superset of FCC-stable structures at the DFT level and predicted using the CV-optimal superset model.}
\end{figure}

\begin{figure}
\centering
\includegraphics[width=0.55\linewidth]{si_fig6_CV-optimal_superset_dravg_model/si_fig6_CV-optimal_superset_dravg_model_v1.pdf}
\caption{\label{fig:figure_superset_dravg_model} Performance metric plots of the CV-optimal superset displacement model evaluated on the superset of FCC-mappable structures. (a) Predicted average displacement vs.\ DFT-level displacement from ideal FCC (b) error of predicted average displacement in histograms binned by elemental fraction of each element, (c) displacement ECIs of clusters plotted vs.\  their respective cluster radii.}
\end{figure}

\newpage

\section{Overview of datasets}

\begin{table}[h]
    \caption{The number of structures in each structure classification subset.\\}
    \centering
    \begin{tabular}{l|r|r|r}
    \hline
               & Loop Data & Auxiliary & Superset \\
 \hline
Relaxed                    & 11,403 & 11,535 &  22,938 \\
FCC-Mappable               & 10,084 &  9,722 &  19,806 \\
Displacement Training Set  &  9,024 & 8,286 & 17,310 \\ 
FCC-Stable                 &  7,708 &  6,581  & 14,289 \\
Energy Training Set        &  5,309 & 4,675 &  9,984 \\ \hline
    \end{tabular}
    \label{tab:dataset_breakdowns}
\end{table}

\newpage

\section{Elemental Fractions and Displacement}

\begin{table}[h]
    \caption{Median of average atomic displacement and (structure counts) by elemental fraction in the FCC-mappable subset of the loop dataset.}
    \footnotesize
	\begin{center}
		\begin{tabular}{|r| m{2.5em} | m{2.5em} | m{2.5em} | m{2.5em} | m{2.5em} | m{2.5em} | m{2.5em} | m{2.5em} | m{2.5em} | m{2.5em} | m{2.5em} | m{2.5em} | m{2.5em} |}
            \hline
			        & \multicolumn{13}{|c|}{Atom Fraction} \\
            \hline
			Element  & 0 & 1/6 & 1/5 & 1/4 & 1/3 & 2/5 & 1/2 & 3/5 & 2/3 & 3/4 & 4/5 & 5/6 & 1 \\
			\hline
			Ni & 0.104 (4095) & 0.103 (2658) & 0.098 (172) & 0.079 (453) & 0.096 (1624) & 0.084 (120) & 0.080 (750) & 0.074 (22) & 0.046 (131) & 0.008 (34) & 0.023 (8) & 0.038 (16) & 0.000 (1) \\
			\hline
			Fe & 0.118 (3871) & 0.105 (2694) & 0.098 (187) & 0.075 (453) & 0.085 (1812) & 0.082 (145) & 0.068 (715) & 0.084 (16) & 0.053 (138) & 0.010 (35) & 0.032 (6) & 0.036 (11) & 0.000 (1) \\
			\hline
			Cr & 0.116 (3641) & 0.096 (2894) & 0.098 (207) & 0.064 (461) & 0.094 (1728) & 0.089 (135) & 0.088 (801) & 0.152 (25) & 0.077 (134) & 0.074 (35) & 0.108 (6) & 0.074 (16) & 0.000 (1) \\
			\hline
			Al & 0.090 (4404) & 0.102 (2715) & 0.093 (167) & 0.089 (455) & 0.111 (1481) & 0.110 (91) & 0.127 (553) & 0.138 (28) & 0.103 (121) & 0.037 (35) & 0.070 (10) & 0.063 (23) & 0.000 (1) \\
			\hline
			Ti & 0.086 (3705) & 0.100 (2728) & 0.093 (181) & 0.078 (438) & 0.108 (1962) & 0.108 (125) & 0.119 (727) & 0.104 (29) & 0.104 (119) & 0.127 (35) & 0.123 (7) & 0.133 (27) & 0.000 (1) \\
			\hline
			Si & 0.084 (4624) & 0.096 (2799) & 0.094 (160) & 0.112 (460) & 0.158 (1393) & 0.144 (76) & 0.237 (432) & 0.180 (17) & 0.305 (66) & 0.537 (34) & 0.311 (7) & 0.562 (15) & 0.000 (1) \\
			\hline
		\end{tabular}
    \label{tab:fracs_and_dravg}
	\end{center}
\end{table}

\newpage

\bibliographystyle{elsarticle-num}
\bibliography{references}